\documentclass{article}
\usepackage{ijcai16}
\usepackage{hyperref}
\usepackage{url}
\usepackage{amsmath}
\usepackage{amssymb}
\usepackage{amsthm}
\usepackage{subfig}
\usepackage{float}
\usepackage{caption}
\usepackage{graphicx}
\usepackage[compact]{titlesec}
\usepackage{times}

\title{HEMI: Hyperedge Majority Influence Maximization \thanks{Copyright {\copyright}  2016 for the individual papers by the papers' authors. Copying permitted for private and academic purposes. This volume is published and copyrighted by its editors.}}

\small

\author{
Varun Gangal, Balaraman Ravindran\\
Department of Computer Science \& Engineering, IIT Madras \\
\texttt{\{vgtomahawk,ravindran.b\}@gmail.com} \\
\AND
Ramasuri Narayanam \\
IBM Research, India \\
\texttt{ramasurn@in.ibm.com} \\
}

\begin{document}

\maketitle
\vspace{-1em}
\begin{abstract}
In this work, we consider the problem of influence maximization on a hypergraph. We first extend the Independent Cascade (IC) model to hypergraphs, and prove that the traditional influence maximization problem remains submodular. We then present a variant of the influence maximization problem (HEMI) where one seeks to maximize the number of hyperedges, a majority of whose nodes are influenced. We prove that HEMI is non-submodular under the diffusion model proposed. 
\end{abstract}

\section{Motivation}
Influence maximization \cite{kempe:03} is a well explored problem in social network analysis. It has widespread applications in political campaigning, viral marketing and understanding the spread of memes and other contagion on social media.

The problem first requires a specification of diffusion model, which specifies how the set of influenced nodes at time $t$ affects the set of non-influenced node at time $t+1$, or in other words, how the influence spreads through the graph $G=(V,E)$. Given a diffusion model, the influence maximization problem aims to find the best initial set of $k$ nodes which have the maximum expected influence at the end of the diffusion process. The function $\sigma(S)$ denotes the expected number of nodes influenced at the end of the diffusion process. 

A hypergraph is a generalization of a graph, where the hyperedges are subsets of vertices (rather than just pairs). More formally, a hypergraph $H=(V,E)$ has a set of vertices $V$, and every $e \in E$ is such that $e \subseteq V$. 

It is not always possible to represent the relationships between actors (nodes/vertices) through the edge, which is a pairwise (dyadic) relation. For instance, in a research setting, researchers collaborate in small groups to write scientific publications. Co-membership of a group in such a setting becomes a super-dyadic relation. A co-authorship network (e.g the arXiv Astrophysics co-authorship network \cite{leskovec2007graph}) can be formed from these collaborations, where every node is a researcher and an edge $(a,b)$ represents $a$ having collaborated with $b$ on a certain publication. However, this representation is lossy, since we lose the information of whether the collaboration of $a$ and $b$ involved a third node $c$. Consider two cases - one where $a$ and $b$, $b$ and $c$, $a$ and $c$ (each pair) worked separately on three publications, and the other where $a$,$b$ and $c$ worked together on one publication. A simple co-authorship network will not be able to distinguish between the two cases. However, a hypergraph representation where every hyperedge is a publication, with its nodes being the researchers who authored the publication, will be able to capture this difference correctly.

\begin{figure}[H]
    \centering
    \includegraphics[scale=0.2]{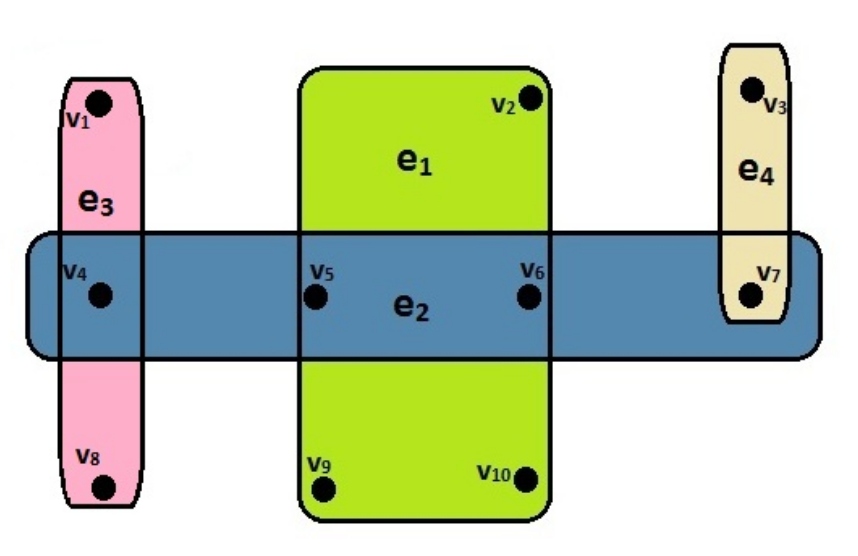}
    \caption{An example hypergraph. $e_1$, $e_2$, $e_3$ and $e_4$ represent the hyperedges}
    \label{fig:exampleHypergraph}
\end{figure}

In the past decade, many standard problems in social network analysis have explored in the context of hypergraphs. \cite{zhou2006learning} looks at spectral clustering and transductive classification with data which can be represented as a hypergraph. \cite{neubauer2009towards} developed algorithms to detect communities in hypergraphs. \cite{satchidanand2015extended} extends random walk based semi-supervised learning to hypergraphs, also handling class imbalance. There are ways of reducing a hypergraph to a simple graph, but none of these are lossless. For example, one method of reducing a hypergraph to a simple graph, known as the clique construction method, adds an edge to a simple graph between $a$ and $b$ when they share atleast one hyperedge. It is however possible to represent a hypergraph as a heterogenous, bipartite graph where one partition consists of the nodes, while the other partition consists of the hyperedges. Conversely, any bipartite graph can be represented as a hypergraph considering one partition to be hyperedges and the other partition to be nodes.  

The problem of influence maximization in a hypergraph has remained hitherto unexplored. \cite{roy2015measuring} is the only work we found which looks at influence maximization in hypergraphs. However, the primary objective of the work is to define Shapley Value based centrality measures for hypergraphs which can be computed on various simple graph reductions constructed from the hypergraph. Influence maximization is just used to evaluate the efficacy of the centrality measures. Moreover, the influence maximization experiments are performed on a clique reduction of a hypergraph rather than the exact hypergraph representation. Hence, the authors simply use existing diffusion models for simple graphs such as IC and LT. Hence, to the best of our knowledge, there are no diffusion models defined which directly model the spread of influence on a hypergraph. Our first motivation, hence is to define a diffusion model for hypergraphs, and analyze whether it is submodular.

In a hypergraph, the nodes represent actors while the hyperedges could be thought of as representing groups or affiliations. At the end of a diffusion process, every hyperedge could have some influenced incident nodes and some non-influenced incident nodes. One may seek to extend the notion of a node being influenced to a affiliation (hyperedge) being influenced. A affiliation is considered influenced if a majority of its nodes (members) are influenced at the end of the diffusion process. There may exist situations where one seeks to maximize the number of affiliations influenced rather than the number of nodes influenced. 

For instance, consider a political campaign on a social network such as Facebook, where the hyperedges correspond to Facebook groups while the nodes correspond to Facebook users. By controlling or influencing a Facebook group, one can dominate the content on the group's page. However, one cannot directly influence an affiliation, since the mechanism of influence propagation takes place through users getting converted in support of the political campaign. The manager of the political campaign here would be interested in selecting the set of initial users (who can be made to support the campaign through promotions or other incentives) which would maximize the number of groups influenced. This is a distinct objective from maximizing the number of nodes influenced, as we shall illustrate now through an example. We refer to this new variant of influence maximization as HEMI (Hyperedge Majority Influence Maximization). $\sigma^{HEMI}(S)$ is the expected number of hyperedges a majority of whose nodes are influenced at the end of the diffusion process.
 
 Consider a hypergraph H where the hyperedges are $e_1=\{v_1,v_2,v_3\}, e_2=\{ v_3,v_4,v_6 \}, e_3 = \{v_3,v_5,v_7\}$. Maximizing the number of nodes would not distinguish between the case where $S_1=\{v_1,v_2,v_3\}$ and $S_2=\{v_3,v_4,v_5\}$ are the final set of nodes influenced, since $\sigma()=3$. However, $\sigma_{HEMI}()=1$ in the first case and $2$ in the second.

An added motivation of solving HEMI rather than maximizing the number of nodes directly would be the diversification it achieves. Attaining diversity for problems such as graph centrality based ranking \cite{mei2010divrank} and k nearest neighbours \cite{ranu2014answering} has been looked at in the past. In an influence maximization setting, one may be interested in having a final set of influenced nodes which is sufficiently diverse. For instance, a cellphone company targeting users to buy its connection, may want the set of users adopting the connection after its first advertising campaign to be well spread through various professions and class verticals. 

\section{Influence Maximization - Preliminaries}
\label{introduction}
The Independent Cascade Model is a well-known diffusion model. Under this model, a node $u$ which is influenced at time $t$ gets one opportunity to influence each of its neighbors $v$ with probability $p_{u,v}$ for time step $t+1$. The probability $p_{u,v}$ is either set to a small constant (e.g $0.1$) or set to $\frac{1}{k_v}$, where $v$ is the target node and $k_v$ is its degree. Another well-known diffusion model is the Linear Threshold (LT) model, where every node $v$ first randomly chooses a threshold $\alpha_v$ $\in [0,1]$. A node $v$ gets influenced at time $t+1$ if more than a fraction $\alpha_v$ of its neighbors are influenced at time $t$.

The major contribution of \cite{kempe:03} was their observation that if the $\sigma()$ function for a diffusion model is both monotone and submodular, the greedy algorithm for growing the candidate set $S$ gives a $(1-\frac{1}{e})$ approximation. Since the brute force algorithm for this problem requires enumerating all the $2^{|V|}-1$ candidate sets, this result provided the first tractable way of solving this problem. The function $\sigma()$ is monotone if $\sigma(S) \leq \sigma(T)$ when $S \subseteq T$. The function $\sigma()$ is submodular if it exhibits a diminishing returns property - in other words for two sets $S$ and $T$ such that $S \subset T$, adding a node $v \notin T$ to S leads to a greater increase in $\sigma()$ than adding $v$ to $T$. 
\begin{equation}
    \sigma(S \cup v) - \sigma(S) \geq \sigma (T \cup v) - \sigma (T)
\end{equation}

The authors then went onto prove that both the IC and LT diffusion models were submodular, i.e they had a submodular $\sigma()$ function. In section \ref{ICProof}, we revisit the proof of submodularity for the IC model, since it is relevant for understanding the proof for submodularity (and counterexample for non-submodularity), which we present later on. 
\section{Proof of submodularity for IC model}
\label{ICProof}
In the IC model, an edge $(u,v)$ is used in the diffusion process with probability $p_{u,v}$. Instead of determining whether $(u,v)$ is used at the time of diffusion, one can perform a trial with respect to each edge $(u,v)$, and form a set of live edges $X \subseteq E$. Only these live edges participate in the diffusion process. We denote by $\sigma_{X}(S)$ the number of nodes influenced given the initial set $S$ and the live edge set $S$. Note that once the live edges are fixed, there is no randomness in the process. Now, one can easily see that $\sigma_{X}(S)$ is simply the size of the set of nodes reachable from $S$ under the graph $G_{X}=(V,X)$. Since reachability is submodular.
\begin{equation}
    \sigma_{X}(S \cup v) - \sigma_{X}(S) \geq \sigma_{X} (T \cup v) - \sigma_{X} (T)
\end{equation}
Taking expectation on both sides, we get 
\begin{equation}
    \sigma(S \cup v) - \sigma(S) \geq \sigma (T \cup v) - \sigma (T)
\end{equation}



\section{Diffusion Model}
Our diffusion model is a simple generalization of the Independent Cascade Model to hypergraphs. Here, we consider a graph $G^{aug}=(V \cup E, E_{aug})$, where the set of nodes $V_{aug}$ is the union of the set of nodes and the set of hyperedges. An edge $e_{aug} = (v,e)$ or $(e,v) \in E_{aug}$ if $v \in V, e \in E v \in e$. For a node-hyperedge pair $(v,e)$ such that $v \in e$, we add edges in both directions ($(e,v)$ and $(v,e)$). This allows us to have independent probabilities of the influence flowing in either direction. In our model, the hyperedges act as carriers of the influence, allowing it to flow from one node to another through them. However, note that in the HEMI objective, we include a hyperedge only if a majority of its nodes are influenced, regardless of whether it has acted as a carrier.

A node $v$ which is influenced at time step $t$ can influence an incident hyperedge $e$ with probability $p_{v,e}$ for time step $t+1$ through the edge $(v,e)$. A hyperedge $e$ which is influenced at time step $t$ can influence an incident node $v$ with probability $p_{e,v}$ for time step $t+1$ through the edge $(e,v)$. Though the probabilities $p_{e,v}$ and $p_{v,e}$ could be set in any way, we use the following two schemes for setting them
\begin{itemize}
    \item $p_{e,v}=p_1$ and $p_{v,e}=p_2$ $\forall v \in V$, $e \in E, v \in e$. Here $p_1$ and $p_2$ are constants. This is similar to the scheme used in  \cite{kempe:03}, where the probability of an edge being live is set to a constant for some of the experiments.
    \item Another scheme would be to normalize all the probabilities of edges incident onto a node in the augmented graph (which could be a node or a hyperedge) to sum to 1. In this case, $p_{e,v}=\frac{1}{k_v}$ and $p_{v,e}=\frac{1}{|e|}$.
\end{itemize}
Here $k_v$ is the degree of the node $v$, while $|e|$ is the size of the hyperedge. 

\begin{figure}[H]
    \centering
    \includegraphics[scale=0.4]{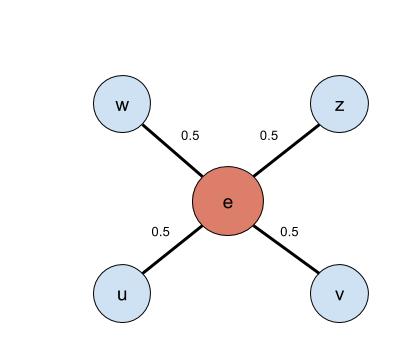}
    \caption{Hypergraph $H$, with a single hyperedge incident on $u$,$v$, $w$, $z$}
    \label{fig:diffusiondepiction}
\end{figure}

Note that our diffusion model does not simplify to a diffusion model on some simple graph reduction of the hypergraph. To understand why this is the case, consider the hypergraph $H$ in Figure \ref{fig:diffusiondepiction} where a hyperedge $e$ has 4 nodes $u$, $v$, $z$ and $w$ incident on it. Also, $u$, $v$, $w$ and $z$ do not have any other hyperedges incident on them. 

Suppose $w$ is already influenced. Now, note that the hyperedge $e$ will get influenced (as a carrier) with probability $0.5$. Let $X_{e}$ denote the 0-1 random variable, which is $1$ if $e$ is influenced, and $0$ otherwise. Similarly, let $X_{u}$, $X_{v}$, $X_{z}$ and $X_{w}$ be the 0-1 random variables corresponding to whether the respective node is influenced or not. $X_{u}$, $X_{v}$ and $X_{z}$ are conditionally independent given $X_e$. We can see this from the fact that $P(X_{u}|X_{e},X_{w}=1)=P(X_{u}|X_{v},X_{e},X_{w}=1)=0.5$. Symmetrically, this holds for the other pairs. 

However, they are not conditionally independent given only $X_w$. We can see that $P(X_{u}|X_{w}=1) = P(X_{u}|X_{e}=1)P(X_{e}=1|X_{w}=1) = 0.5 \times 0.5 = 0.25$. 
Let us now find the value of $P(X_{u}|X_{w}=1,X_{v}=1)$.

\tiny
\begin{align}
    P(X_{u}=1|X_{w}=1,X_{v}=1) &= P(X_{u}=1|X_{e}=1)P(X_e=1|X_w=1,X_v=1) \\
                             &= 0.5\frac{P(X_v=1|X_e=1)P(X_e=1|X_w=1)}{P(X_v=1|X_w=1)} \\
                             &= 0.5 \frac{P(X_v=1|X_e=1)P(X_e=1|X_w=1)}{P(X_v=1|X_w=1)} \\
                             &=0.5 \\
\end{align}
\normalsize
Since $P(X_{u}|X_{w}=1,X_{v}=1) > P(X_{u}|X_{w}=1)$, we can see that $X_{u}$ and $X_{v}$ are not conditionally independent given only $X_{w}$. A simple graph based representation, with the IC diffusion model  will not be able to model the dependence between $X_{u}$ and $X_{v}$, given $X_{w}$ at time $t$. This is because the state of $X_{u}$ and $X_{v}$ at the next time step (t+1), only depends on whether the respective edges $(w,u)$ and $(w,v)$ become live. The triadic relation captured by the hyperedge between $w$, $u$ and $v$ is not captured here. \cite{agarwal2006higher} had shown that most of the hypergraph based Laplacians were reducible to the Laplacians of two simple graph constructions - the star expansion and the clique expansion. However, for our diffusion model, we have shown that it is necessary to work with the exact hypergraph representation.

\section{Proof of submodularity}
We shall first prove that the function $\sigma(S)$ where $S$ is the number of initially influenced nodes and $\sigma(S)$ is the final number of nodes influenced, is submodular under our diffusion model. Our proof is a simple generalization of the proof for submodularity of Independent Cascade in simple graphs.

Consider the graph $G^{aug}$. Assume, as in the original IC proof, that the set of live edges $X$ is decided by performing all the trials prior to the diffusion process. Consider sets of nodes in the original hypergraph $S \subseteq V, T \subseteq V, S \subset T$, and a node in the original hypergraph $v \notin T$. Now, $\sigma(S)$ is the set of augmented graph nodes which were nodes in the original graph and are reachable from the set $S$ in the augmented graph. Note that this does not include the augmented graph nodes which correspond to hyperedges, though the paths which are used to reach a node $v \in V$ from $u \in S$ may have hyperedges as intermediate nodes. 

Using the same argument as in the KKT proof, since reachability is submodular, we get
\begin{equation}
    \sigma_{X}(S \cup v) - \sigma_{X}(S) \geq \sigma_{X} (T \cup v) - \sigma_{X} (T)
\end{equation}
Taking the expectation over all X,
\begin{equation}
    \sigma(S \cup v) - \sigma(S) \geq \sigma (T \cup v) - \sigma (T)
\end{equation}

\section{The HEMI Objective - Formal Definition}
Consider the state of the hypergraph $H=(V,E)$ at the end of a trial of the diffusion process, for a given initial set $S$. Some of the nodes will be influenced, and some will not be influenced. Let the set of influenced nodes be $I$. The set of non-influenced nodes will be $V-I$. Clearly $\sigma(S)$, the expected number of nodes influenced, is equal to $E[|I|]$.

Let $V_{e}$ denote the set of nodes incident on a given hyperedge $e$. For HEMI, we are interested in the number of hyperedges, a majority of whose nodes are influenced. We define $Y$ to be this set. More formally,
\begin{equation}
    Y = \{e \in E| V_e \cap I \geq \frac{|V_e|}{2}+1 \}
\end{equation}
Now, $\sigma^{HEMI}(S) = E[|Y|]$.

\section{Non-submodularity of HEMI:Counterexample}
\begin{figure}[H]
    \centering
    \includegraphics[scale=0.4]{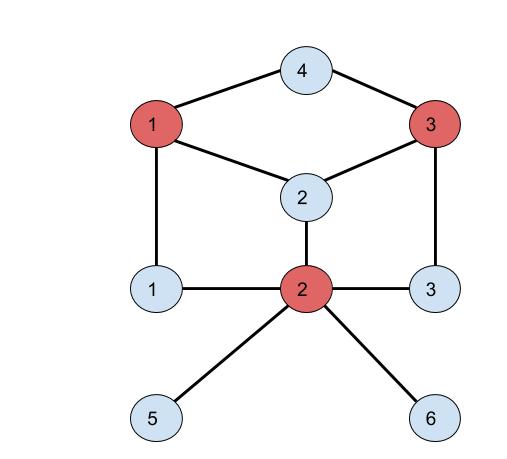}
    \caption{Hypergraph $H$; the red ovals correspond to hyperedges while the blue ovals correspond to nodes.}
    \label{fig:HEMICounter}
\end{figure}

Consider the hypergraph $H$ shown in Figure \ref{fig:HEMICounter}. $H$ has hyperedges $e_1 = \{v_1, v_2, v_4\}$, $e_2=\{v_1, v_2, v_3, v_5,v_6\}$, $e_3= \{v_2,v_3,v_4\}$. 
The augmented graph $G^{aug}$ will have $2\times(3+5+3)=22$ edges.  Let $\sigma^{HEMI}(S)$ be the number of hyperedges with a majority of their nodes influenced at the end of the diffusion process. Let $\sigma^{HEMI}_{X}(S)$ be the number of majority-influenced hyperedges given a set of live edges $X$ in the augmented graph. One can exactly compute $\sigma(S)$ by enumerating over all possible live sets $X$. 
\begin{equation}
    \sigma^{HEMI}(S) = \sum_{X} P(X) \sigma^{HEMI}_{X}(S) 
\end{equation}
where P(X) is given by
\begin{equation}
    P(X) = \Pi_{e_{aug} \in X} p(e_{aug})
\end{equation}
Here $p(e_{aug})=p_{e,v}$ or $p_{v,e}$ depending on the directionality of the edge i.e whether it goes from a node to a hyperedge or in the reverse direction. However, note that using this method of enumerating outcomes even for this small example, would require us to enumerate $2^{22}$ outcomes. Instead, we find an estimate of $\sigma^{HEMI}(S)$ using a simulation-based method. 

Consider the sets $S=\{v_5\}$, $T=\{v_1,v_3,v_5\}$ and $v=v_2$. Now, we run simulations to estimate $\sigma^{H}(S)$, $\sigma^{H}(S \cup v)$, $\sigma^{H}(S \cup v)$ and $\sigma^{H}(T \cup v)$. For a given set, we run $10^5$ trials. In each trial, we start with the initial set influenced at time step $0$, and then run the diffusion model until convergence (set of influenced nodes stops growing). We then find the number of hyperedges a majority of whose nodes are influenced. We find this number averaged over all the trials. 

We can see that submodularity is violated for a wide range of $p_1$ and $p_2$ values. The intuition here is that node $v_2$, which is incident on all the hyperedges, can act as a swing vertex (converting a non-majority into a majority) only when one less than majority of the nodes in a hyperedge it belongs to are influenced. A larger initial set helps having more influenced nodes in the end, thus helping $v$ perform the role of converting a non-majority into a majority in more hyperedges. Hence, a larger initial set results in a larger marginal gain on adding $v$ to the initial set.
\begin{table}
\centering
\begin{center}
\begin{tabular}{ | c | c | c | c |}
\hline \hline
$p_{v,e}$ & $p_{v,e}$ & $\sigma^H(S \cup v) - \sigma^H(S)$ & $\sigma^H(T \cup v) - \sigma^H(T)$ \\ \hline \hline
0.1 & 0.1 & 0.08 & 1.92 \\ \hline 
0.1 & 0.2 & 0.13 & 1.85 \\ \hline
0.1 & 0.3 & 0.17 & 1.79 \\ \hline
0.2 & 0.1 & 0.15 & 1.85 \\ \hline
0.2 & 0.2 & 0.26 & 1.71 \\ \hline
0.2 & 0.3 & 0.33 & 1.60 \\ \hline
0.3 & 0.1 & 0.22 & 1.78 \\ \hline
0.3 & 0.2 & 0.38 & 1.60 \\ \hline
0.3 & 0.3 & 0.48 & 1.43 \\ \hline
$\frac{1}{|e|}$ & $\frac{1}{d(v)}$ & 0.80 & 0.85 \\ \hline
\end{tabular}
\end{center}
\end{table}

Thus, we can see that the HEMI objective $\sigma^{HEMI}(S)$ is non-submodular. This means the greedy algorithm is no longer guaranteed to provide a $(1-\frac{1}{e})$ approximation. 

\section{Future Work}
Although we have proved that the HEMI objective is non-submodular, we are yet to devise an algorithm which can solve the problem with some provable guarantee. Devising such an algorithm is a point of future interest. One possible direction could be to formulate the HEMI objective as a difference of two submodular functions, or having submodular upper and lower bounds.
\bibliographystyle{named}
\bibliography{ijcai16}

\end{document}